\documentclass[aps,prb,superscriptaddress,twocolumn,showpacs,10pt]{revtex4-1}
\usepackage{amsmath,amssymb,mathrsfs,bbold}
\usepackage[pdftex]{graphicx, color}
\usepackage{epstopdf}

\allowdisplaybreaks[4]

\begin{document}
\title{Critical scales in anisotropic spin systems from functional renormalization}
\author{Stefan G\"ottel}
\affiliation{Institute for Theory of Statistical Physics, RWTH Aachen,
52056 Aachen, Germany}
\affiliation{JARA-Fundamentals of Future Information Technology}
\author{Sabine Andergassen}
\affiliation{Institute for Theory of Statistical Physics, RWTH Aachen,
52056 Aachen, Germany}
\affiliation{JARA-Fundamentals of Future Information Technology}
\affiliation{Faculty of Physics, University of Vienna, Boltzmanngasse 5, 1090 Vienna, Austria}
\author{Carsten Honerkamp}
\affiliation{JARA-Fundamentals of Future Information Technology}
\affiliation{Institute for Theoretical Solid State Physics, RWTH
Aachen, 52056 Aachen, Germany}
\author{Dirk Schuricht}
\affiliation{Institute for Theory of Statistical Physics, RWTH Aachen,
52056 Aachen, Germany}
\affiliation{JARA-Fundamentals of Future Information Technology}
\author{Stefan Wessel}
\affiliation{JARA-Fundamentals of Future Information Technology}
\affiliation{Institute for Theoretical Solid State Physics, RWTH
Aachen, 52056 Aachen, Germany}
\affiliation{JARA-High-Performance Computing}

%

\date{\today}
\pagestyle{plain}

\begin{abstract}
We apply a recently developed functional renormalization group (fRG) scheme for quantum spin systems to the spin-1/2 antiferromagnetic XXZ model on a two-dimensional square lattice.
Based on an auxiliary fermion representation we derive flow equations which allow a resummation of the perturbation series in the spin-spin interactions.
Spin susceptibilities are calculated for different values of the anisotropy parameter. The phase transition between planar and axial ordering at the isotropic point is reproduced correctly.
The results for the critical scales from the fRG as quantitative measures for the ordering temperatures are in good agreement with the exact solution only in the Ising limit. 
In particular on the easy-plane side, the deviations from critical temperatures obtained with quantum Monte Carlo are rather large.
Furthermore, at the isotropic point the Mermin-Wagner theorem is violated such that a description of the critical behavior and an extraction of scaling exponents is not possible. 
We discuss possible reasons for these discrepancies.
\end{abstract}

\maketitle

\section{Introduction}
Renormalization group (RG) methods for fermions developed over the last decades have become a widely used tool for correlated electron systems\cite{solyom,shankarRMP,metznercastellani,metznerRMP}.
Besides a qualitative understanding of fundamental low-energy aspects of weakly interacting many-fermion systems, such as the stability of Fermi liquids\cite{shankarRMP,salmhoferbook} 
or the effective behavior of impurities in Luttinger liquids\cite{andergassen04,andergassen04b}, RG methods have also been used extensively for investigating competing ordering tendencies in 
two-dimensional lattice systems or the different regimes in impurity problems in one or zero dimensions\cite{metznerRMP,andergassen_review}. 
Functional RG (fRG) methods do not only concentrate on a smaller number of possibly relevant terms in the theory, but aim to resolve the flow of the generating functional of the theory 
or the respective vertex functions in as much detail as possible. This allows one to monitor how effects at intermediate energy scales drive the flow to certain fixed points, 
and at which scales typical low-energy behaviors such as power-laws or scaling should actually hold. While for bosonic problems one can implement the fRG non-perturbatively\cite{berges,kopietz}, 
for fermionic problems\cite{metznerRMP,kopietz} one has to work with a truncation in the powers of fermionic fields. This usually restricts the validity to weak to moderate interaction strengths. 
Many applications use the fRG method primarily as a qualitative tool for the exploration of theoretical pathways to different low-energy behaviors. 
Nevertheless, it is important to note that for most cases alluded to above, the approach is controllable in the limit of small interaction strengths. 
For example, the exponents in Luttinger liquids obtained using the fRG are consistent with non-perturbative or exact solutions\cite{andergassen04,schuetz05}. 
Likewise, the $d$-wave superconducting pairing instability in the two-dimensional Hubbard model can be seen to be immune against corrections in the limit of small interactions\cite{raghu}.

Competing orders and strong quantum fluctuations are also encountered in quantum spin systems\cite{qspinreview,qspinreview2}. In this very active research area, analytical or field theoretical approaches 
are, however, even more involved than for moderately interacting fermions due to the constraint of a fixed total spin per lattice site. Direct numerical methods are often 
the most reliable approach to these strongly correlated problems, but finite-size effects and sign problems due to frustration effects can make it difficult to obtain clear results. 
Hence new methods might help. Recently, the fRG has been formulated for auxiliary $S=1/2$ fermions describing localized quantum spins. 
In a series of papers\cite{ReutherWoelfle10,Reuther-10a,Reuther-11,Reuther-11a,Reuther-11b,Schmidt-10,Singh-12}  it was shown that this approach gives qualitatively appealing results, 
e.g., for frustrated systems on various two-dimensional lattices. As known from other problems, one advantage of the fRG with respect to other numerical methods is the flexibility 
in treating different lattice geometries and dimensions. With manageable numerical effort one can study rather large, frustrated systems without encountering sign problems. 
The formulation is transparent; for example, competing trends can be identified and manipulated (by modifying terms) in the calculations. 
Hence it is an interesting question how far the  spin-fRG approach can be pushed and how it compares quantitatively to other methods, in cases where benchmarking is possible.

Testing the spin-fRG is also interesting from a theoretical point of view, for several reasons. Expressing the localized spins in terms of auxiliary fermions brings some important differences 
with respect to 'normal' fermions on a lattice.
(i) As mentioned above, the auxiliary fermion number on each site should remain constrained to a fixed value in order to guarantee 
a faithful representation of the localized quantum spins. Most of the previous applications of the spin-fRG used a mean-field-like approximation at $T=0$ where the constraint is treated on average. 
We will adopt this approximation here as well. In Ref.~\onlinecite{ReutherWoelfle10} the authors also report on the possibility to go beyond this approximation by using 
an imaginary chemical potential\cite{Popov-88} at nonzero temperature $T$. However, this implies a substantial increase in the numerical effort and we do not consider this option here. 
(ii) The bare action in terms of the auxiliary fermions does not contain a kinetic energy, i.e., the auxiliary fermions are dispersionless.  
This is, of course, a prerequisite for the particle-number constraint. Fortunately, in the fRG for this system the dispersionless character is preserved and the auxiliary fermion self-energy 
remains local, which also significantly simplifies the numerical treatment. On the other hand, the absence of a kinetic energy means that, at least at $T\to 0$, the system is strongly coupled, 
and {\it a priori} there is no reason why a perturbative fermionic fRG setup (or any other theory perturbative in the interactions) should work well. 
(iii) Quadratic interactions between localized spins translate into quartic auxiliary fermion interactions that are bi-local in the lattice index, i.e., consist of two pairs of fermion fields 
with the same site index. Again, this structure is preserved under the fRG flow. This is a tremendous simplification with respect to the 'normal' fermionic case where in general 
also interactions with more than two site indices occur. 
Concluding this discussion of the auxiliary fermionic description one can say that points (i) and (ii) are at least two major uncertainties in the spin-fRG approach. 
On the positive side, the simplified structure of both the bare and the flowing theory allows one to capture basically the full frequency dependence of the self-energy 
and the four-point interaction vertices, and in addition makes it possible to include the self-energy feedback in the flow equations without too much effort. 
This means that in the spin-fRG, one reaches an advanced approximation level (for extended systems in more than one dimension) that has before only been used for impurity systems in very selected 
cases\cite{karrasch,Jakobs-10}. Hence, it is a very interesting question how this combination of positive and negative expectations performs if one tries to compare the results with other techniques.

The goal of the present paper is to provide a quantitative test of the spin-fRG.
In order to provide a well-understood testing ground, we study the spin-$1/2$ antiferromagnetic 
XXZ-model, which as function of the anisotropy parameter is known to exhibit two different 
orderings, and where the critical temperatures for these orderings show a nontrivial variation.
The critical scales $\Lambda_c$ for these orderings computed with the spin-fRG are compared 
with the critical temperatures $T_c$  obtained from quantum Monte Carlo (QMC) calculations.
We note that while it is in principle possible to implement the spin-fRG for finite temperatures
and hence to directly measure $T_c$, this would drastically increase the numerical effort.
Thus in practice one aims to relate the energy scales $\Lambda_c$ found more easily 
at $T=0$ to the actual $T_c$. Our work provides insights to where and to what extent this is
possible.

This paper is organized as follows: after introducing the anisotropic Heisenberg model, 
the auxiliary-fermion representation and the technical implementation of the fRG specific to spin systems is shortly reviewed. 
We then present the results for the spin susceptibility and the critical scales of the fRG as estimates for the critical temperature and discuss the merits and limitations of the spin-fRG.

\section{Model} \label{sec:model}
In the following, we consider the spin-1/2 XXZ model with antiferromagnetic nearest-neighbor exchange interaction $J > 0$, defined by the Hamiltonian
\begin{equation}
 H = J \sum_{\langle i j \rangle} \left( S^x_i S^x_j + S^y_i S^y_j + \Delta S^z_i S^z_j \right).
\end{equation}
Here, $\vec{S}_i$ denotes the spin-1/2 operators on lattice site $i$ of a two-dimensional square lattice with periodic boundary conditions.
Furthermore, $\Delta$ is the exchange anisotropy, which we restrict to the antiferromagnetic domain, i.e., $\Delta \geq 0$, throughout this study. 
For the above model, both the ground state properties as well as the finite temperature phase diagram are well established, cf., e.g., Ref.~\onlinecite{Farnell-04}. 
Here, we mention those features that are relevant for our assessment of the spin-fRG method.

In the easy-axis region, $\Delta>1$, the ground state is N\'eel-ordered along the longitudinal ($z$) direction, while within the easy-plane region for $\Delta<1$, 
the ground state exhibits long-range transverse antiferromagnetic ordering within the $xy$-plane. Within the easy-axis region, long-ranged longitudinal N\'eel-order emerges already 
below a finite transition temperature $T_c>0$. This  transition corresponds to the breaking of the discrete $Z_2$ symmetry of $H$, and thus belongs to the universality class of the two-dimensional 
Ising model. In particular, in the large-$\Delta$ limit, $\Delta \rightarrow \infty$, the transition temperature scales like $T_c\approx 2.269 \Delta J /4$, as obtained from the exact solution of 
the two-dimensional Ising model~\cite{Onsager-44} (the factor of $1/4$ results from the different normalization of the spin variables usually employed in the classical Ising model as compared 
to the spin-1/2 operators entering $H$). The ground state ordering in the easy-plane region, i.e., for $\Delta<1$, relates to the breaking of the $U(1)$ symmetry of $H$, such that in two dimensions 
the transverse long-range order does not persist to nonzero temperatures, consistent with the Mermin-Wagner theorem~\cite{Mermin-66}. Still, the system exhibits a Berezinskii-Kosterlitz-Thouless (BKT) 
transition~\cite{berezinskii,kosterlitz} at a critical temperature $T_{\mathrm{BKT}}>0$ driven by the unbinding of vortex excitations above $T_{\mathrm{BKT}}$ out of a low-temperature 
quasi-long-ranged ordered phase with algebraic decay of the transverse spin correlation function. This phase is furthermore characterized by a finite spin stiffness, which exhibits an universal jump 
at $T_{\mathrm{BKT}}$. At $\Delta = 1 $, the Hamiltonian $H$ exhibits the full $SU(2)$ symmetry and the antiferromagnetic order is constrained to zero temperatures, 
such that at this isotropic point the critical temperature vanishes, i.e., $T_c=0$. The $\Delta$-dependence of both the Ising transition temperature and the BKT transition temperature 
in the vicinity of $\Delta=1$ has been analyzed in Ref.~\onlinecite{Cuccoli-03}. 

However, in order to assess the quality of the spin-fRG approach to the XXZ model, we require an estimation of the critical temperature $T_c$ for a wider range of different $\Delta$-values. 
Since the model is not frustrated, we can indeed employ numerically exact, unbiased QMC methods for this purpose. Here, the critical temperatures in both the easy-axis and easy-plane regions 
have been extracted from QMC simulations using the stochastic series expansion method~\cite{sse1} with generalized directed loop updates~\cite{sse2,sse3}. In particular, 
we simulated finite lattices with $N=L^2$ spins and with linear system sizes $L$ ranging up to $32$ using periodic boundary conditions. From the finite size data, the critical temperature for the 
Ising transition was obtained from a finite-size analysis based on the critical Binder ratio~\cite{binder}. Within the easy-axis region, the spin stiffness
was calculated from measuring the spin winding number fluctuations~\cite{pollock}. From the universal stiffness jump at the BKT transition, the transition temperature was then obtained~\cite{bkt}.
The values of the transition temperatures for $|\Delta-1|<0.1$ can also be taken from Ref.~\onlinecite{Cuccoli-03} and for $\Delta = 0$ from Ref.~\onlinecite{Harada-97}. 
The results of these calculations will be presented below. However, before presenting these data, we provide a short introduction to the spin-fRG method in the following section.

\section{Auxiliary fermions and spin-fRG}
Here we briefly describe the spin-fRG scheme introduced by Reuther and 
W\"olfle~\cite{ReutherWoelfle10}.
In order to be able to implement the fermionic fRG scheme\cite{salmhofer01,metznerRMP}  for the quantum spin system, one chooses a representation of the spin operators in terms of 
Abrikosov auxiliary-fermions \cite{Abrikosov-65}. This representation requires two fermionic operators $f_{i \alpha}$, $\alpha = \uparrow, \downarrow$, for each lattice site $i$,
\begin{equation}
 S^\mu_i = \frac 12 \sum_{\alpha \beta} f^\dagger_{i \alpha} \sigma^\mu_{\alpha \beta} f_{i \beta},
\end{equation}
with $\sigma^\mu$ being the Pauli matrices. By construction, the representation satisfies the correct commutation relation of spin operators. The advantage of this representation lies in
its quadratic form which allows for the application of Feynman-diagram techniques. However, the introduction of auxiliary fermions implies an enlargement of the Hilbert space, 
as a single site can now also carry total spin zero, either by being empty or by being doubly occupied. In order to obtain a faithful representation of the quantum spin-$1/2$-system, these unphysical 
states have to be projected out. This is formally achieved by the requirement that the single-occupancy operators 
\begin{equation}
 Q_i = \sum_{\alpha} f^\dagger_{i \alpha} f_{i \alpha}  \;
\end{equation}
all possess the eigenvalue 1 on any physical state $| \mathrm{phys} \rangle$ occurring in the thermodynamic averages of the fermionic system, i.e. 
$Q_i | \mathrm{phys} \rangle = | \mathrm{phys} \rangle$. Thus all unphysical states are discarded.

The single-occupancy constraint immensely complicates the fRG treatment. 
A convenient approximation is to enforce the constraint only on average by requiring 
$\langle Q_i \rangle =1$. With this, of course, local particle number fluctuations 
are still possible.
Due to the particle-hole symmetry in the dispersionless pseudo-fermion system 
$\langle Q_i \rangle =1$ is fulfilled for $\mu=0$. 
For an exact projection without fluctuations, Popov and Fedotov \cite{Popov-88} proved the 
mutual cancellation of the unphysical contributions at each lattice site in the thermodynamic 
averages if an imaginary chemical potential $\mu = - i \pi T/2$ is used. However, as the 
limit $T\to 0$ and the suppression of fluctuations at $T=0$  do not necessarily commute, the 
two procedures need not to be equivalent at $T=0$. 
Following Ref. \onlinecite{ReutherWoelfle10} we here consider the average projection  $\langle Q_i \rangle =1$ and vanishing temperature $T=0$ for simplicity. 
More details to this point can be found in Ref. \onlinecite{Reuther-PhD}.

After transforming the initial spin-system into a fermionic one, the requirements to apply the fermionic fRG are fulfilled (except for the lack of a small expansion parameter; see the 
discussion below).
The idea\cite{metznerRMP} is to sum up the contributions to the one-particle irreducible (1PI) 
vertex functions from high to low energies. To achieve this,
we introduce a cutoff $\Lambda$ into the bare Green's function which is replaced by
\begin{equation}
 G^{0,\Lambda}(i \omega) = \frac{\Theta(|\omega| - \Lambda)}{i \omega}\;.
\end{equation}
Inserting $G^{0,\Lambda}$ into the generating functional of the 1PI vertices and performing the derivative with respect to $\Lambda$ 
leads to an infinite hierarchy of flow equations.
For numerical calculations this hierarchy has to be truncated. It turns out that a truncation after the four-point vertex in the straightforward form of the 1PI fRG 
hierarchy\cite{salmhofer01} is not sufficient to adequately describe the competition between order and disorder fluctuations\cite{ReutherWoelfle10}. 
An improved scheme suggested by Katanin \cite{Katanin-04} includes more self-energy corrections such that Ward identities \cite{Ward-50,Enss-05} are better fulfilled. 
With this Katanin-modification, which we will use as well, the results for the $J_1$-$J_2$-model were found\cite{ReutherWoelfle10} to be in good agreement with other studies.
However, we will see that the spin fRG still overestimates the magnetic ordering tendencies during the flow.
Unfortunately, due to the increasing numerical effort the inclusion of higher-order contributions beyond the Katanin truncation scheme is not feasible.

Within the Katanin-modified 1PI-scheme, the RG equations for the self-energy 
$\Sigma^\Lambda$ and the two-particle vertex $\Gamma^\Lambda$ read
\begin{eqnarray}
\frac{d}{d\Lambda}\Sigma^\Lambda_1 &=& - \frac{1}{2 \pi} \sum_2 \Gamma^\Lambda_{1,2;1,2} S^\Lambda(\omega_2) \label{eq::FlowSigma} \\
\frac{d}{d\Lambda}\Gamma^\Lambda_{1',2';1,2} &=&- \frac{1}{2 \pi} \sum_{3,4} \big[ \Gamma^\Lambda_{1',2';3,4} \Gamma^\Lambda_{3,4,1,2} \notag\\*
&&-\Gamma^\Lambda_{1',4;1,3} \Gamma^\Lambda_{3,2';4,2} - (3 \leftrightarrow 4) \notag\\*
&& + \Gamma^\Lambda_{2',4;1,3} \Gamma^\Lambda_{3,1';4,2} + (3 \leftrightarrow 4) \big]  \notag\\*
&&\times G^\Lambda(\omega_3) \frac{d}{d\Lambda}G^\Lambda(\omega_4) \label{eq::FlowGamma}
\end{eqnarray}
where the multi-indices $1 = {\omega_1, i_1, \alpha_1}$ include the frequency, site, and spin, and the sum over $1$ contains an integral over $\omega_1$ as well as a sum over $i_1$ and $\alpha_1$. 
Furthermore, 
\begin{equation}
G^\Lambda(i \omega) = \frac{\Theta(|\omega| - \Lambda)}{i \omega - \Sigma^\Lambda(i \omega)}
\end{equation}
is the full propagator and
\begin{equation}
S^\Lambda(i \omega) = \frac{\delta(|\omega| - \Lambda)}{i \omega - \Sigma^\Lambda(i \omega)}
\end{equation}
denotes the single-scale propagator.
The self-energy is a purely imaginary odd function of the frequency and can be written in the form
\begin{equation}
\Sigma^\Lambda_i(i \omega) = - i \gamma^\Lambda(\omega) \label{eq::ZerlegungSigma}. 
\end{equation}
We observe that in the diagrammatic expansion the Green's function remains strictly local\cite{ReutherWoelfle10} and the momentum dependence of the susceptibility is generated by the non-local 
exchange couplings. If one uses the symmetry relations of the system and energy conservation, one can parametrize the two-particle vertex in the form
\begin{eqnarray}
\Gamma^\Lambda_{1', 2'; 1, 2} &=& \Big[ \Gamma_{z i_1 i_2}^\Lambda(s, t, u) \sigma^z_{\alpha_{1'} \alpha_1} \sigma^z_{\alpha_{2'} \alpha_2} \notag\\*
&& + \Gamma_{xy i_1 i_2}^\Lambda(s, t, u) (\sigma^x_{\alpha_{1'} \alpha_1} \sigma^x_{\alpha_{2'} \alpha_2} 
+ \sigma^y_{\alpha_{1'} \alpha_1} \sigma^y_{\alpha_{2'} \alpha_2}) \notag\\*
&& +  \Gamma_{d i_1 i_2}^\Lambda(s, t, u) \delta_{\alpha_{1'} \alpha_1} \delta_{\alpha_{2'} \alpha_2} \Big] \times \delta_{i_{1'} i_1} \delta_{i_{2'} i_2} \notag\\*
&& -\Big[\Gamma_{z i_1 i_2}^\Lambda(s, u, t) \sigma^z_{\alpha_{1'} \alpha_2} \sigma^z_{\alpha_{2'} \alpha_1} \notag\\*
&&+ \Gamma_{xy i_1 i_2}^\Lambda(s, u, t) (\sigma^x_{\alpha_{1'} \alpha_2} \sigma^x_{\alpha_{2'} \alpha_1}
+ \sigma^y_{\alpha_{1'} \alpha_2} \sigma^y_{\alpha_{2'} \alpha_1}) \notag\\*
&& +  \Gamma_{d i_1 i_2}^\Lambda(s, u, t) \delta_{\alpha_{1'} \alpha_2} \delta_{\alpha_{2'} \alpha_1} \Big] \times \delta_{i_{1'} i_2} \delta_{i_{2'} i_1} \label{eq::ZerlegungGamma}
\end{eqnarray}
with 'Mandelstam'-variables $s = \omega_1 + \omega_2$, $t = \omega_{1'} - \omega_1$ and $u = \omega_1 - \omega_{2'}$. To be more precise, the vertices depend only on the difference $|i_1 - i_2|$ 
due to translational invariance.
The resulting RG equations for $\gamma(\omega)$, $\Gamma_{zij}(s, t, u)$, $\Gamma_{xyij} (s,t,u)$, and $\Gamma_{dij}(s,t,u)$ including the full frequency dependence are explicitly 
stated in Ref. \onlinecite{Goettel-11}.
The initial conditions are
\begin{eqnarray}
 \gamma^{\Lambda = \infty} &=& 0,\label{eq:init1} \\
 \Gamma^{\Lambda = \infty}_{xyi_1 i_2}(s, t, u) &=& \begin{cases}
                                                  \frac{1}{4} J \qquad \textnormal{ for } \quad |i_1 - i_2| = 1,\\
						  0 \qquad \qquad \textnormal { otherwise},
                                                 \end{cases}\\
 \Gamma^{\Lambda = \infty}_{zi_1 i_2}(s, t, u) &=& \begin{cases}
                                                  \frac{\Delta}{4} J \qquad \textnormal{ for } \quad |i_1 - i_2| = 1,\\
						  0 \qquad \qquad \textnormal { otherwise},
                                                 \end{cases}\\
 \Gamma^{\Lambda = \infty}_{di_1 i_2}(s, t, u) &=& 0.\label{eq:init4}
\end{eqnarray}
We note that the spatial structure of the interactions only enters the initial conditions.

We can also drop some terms in this 'full' set of fRG equations and thereby reproduce single-channel summations of random-phase-approximation (RPA) character. Below we will 
present RPA data from such a restricted flow, where only specific terms in the $t$-channel 
[$\sim \Gamma^\Lambda_{1',4;1,3} \Gamma^\Lambda_{3,2';4,2} + (3 \leftrightarrow 4)$] 
of the flow equation (\ref{eq::FlowGamma}) for the two-particle vertex are taken into account, as explained in Ref. \onlinecite{ReutherWoelfle10}. 
Inserting the representation (\ref{eq::ZerlegungGamma}) into the $t$-channel gives various terms. The terms involving $\Gamma_{di_1i_2}^\Lambda$ remain zero, 
as the initial condition $\Gamma_{di_1i_2}^{\Lambda=\infty}=0$ is conserved in the flow with the $t$-channel only. The remaining terms can be analyzed most easily by Fourier-transforming 
the vertices $\Gamma^{\Lambda}_{z/xy \, i_1i_2}$ from $i_1 - i_2$ to $\vec{q}$. In reciprocal space, the vertices only depend on one wave-vector $\vec{q}$. 
In the $t$-channel for the flow of $\Gamma^{\Lambda}_{z/xy \, \vec{q}}$  there are now terms that do not couple different $\vec{q}$s, while other terms contain an internal sum over another 
wave-vector $\vec{q}'$. The first class of terms can be interpreted as bubble-like RPA, while the second class could be called vertex corrections, and are dropped in the RPA-approximation. 
This leads to the RPA-like flow equation for the vertex, explicitly written out in Eq. (40) in Ref. \onlinecite{ReutherWoelfle10}. 
We discuss two versions: In the ``RPA0''  the self-energy is neglected, whereas for the ``RPA+'' self-energy contributions are taken into account as in the 
usual self-consistent RPA without cutoffs. For the flow equation of the self-energy no further selection of the diagrams is necessary, here only a Fock-like contribution remains. 
Note that in specific fermionic models, integrating the RPA + Hartree terms gives results equivalent to mean-field theory\cite{Katanin-04,gapflow}, 
but RPA + Fock (or, equivalently, mean-field+Fock, as the Hartree term vanishes) as in our case does not allow such an equivalence. Furthermore, note that the self-consistent RPA for the 2D Heisenberg-antiferromagnet containing the same class of 
RPA + Fock diagrams has been found to prevent long-range order at $T>0$ in accordance with the Mermin-Wagner theorem, but at the price of overdamping the spin-wave excitations 
at small frequencies\cite{brinckmann}. When comparing this to the RPA+, it is important to note that while diagrams with the same topology are used, the propagators on the internal lines differ, 
and in general the fRG results integrated down to zero scale cannot be expected to match those of self-consistent schemes.

\section{Results}
We have solved the RG equations \eqref{eq::FlowSigma} and \eqref{eq::FlowGamma}
with the initial conditions \eqref{eq:init1}--\eqref{eq:init4} numerically. 
Following Ref.~\onlinecite{ReutherWoelfle10} we used a logarithmic 
frequency discretization of $N_f = 30$ positive frequencies, if it is not explicitly stated otherwise, with $\omega_{min} = 0.001J$ and $\omega_{max} \approx 250J$. 
This ensures that the small frequencies are covered in detail and some frequencies higher than $\Delta \times J$ are used. 
Negative frequencies are implicitly included via symmetry relations. Unless stated otherwise, all results were obtained
for a lattice with $10 \times 10$ sites. Different from the original treatment by Reuther and W\"olfle\cite{ReutherWoelfle10}, the interactions are not truncated in range. Instead periodic boundary conditions are used. 

\subsection{Susceptibility}
First we compute the static longitudinal spin-spin correlation function $\chi_{ij}^{zz}$ by a diagrammatic calculation 
\begin{eqnarray}
\chi_{ij}^{zz}(i \nu = 0) &&= \int_0^\infty d \tau \left< T_\tau \left[ S_i^z(\tau) S_j^z(0) \right] \right> \\
\notag\\
&& \begin{picture}(0,0)%
\includegraphics{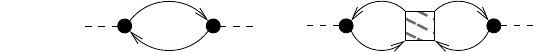}%
\end{picture}%
\setlength{\unitlength}{2072sp}%
\begingroup\makeatletter\ifx\SetFigFont\undefined%
\gdef\SetFigFont#1#2#3#4#5{%
  \reset@font\fontsize{#1}{#2pt}%
  \fontfamily{#3}\fontseries{#4}\fontshape{#5}%
  \selectfont}%
\fi\endgroup%
\begin{picture}(4887,476)(166,-7232)
\put(2656,-7081){\makebox(0,0)[lb]{\smash{{\SetFigFont{8}{9.6}{\familydefault}{\mddefault}{\updefault}{\color[rgb]{0,0,0}+}%
}}}}
\put(1396,-7037){\makebox(0,0)[lb]{\smash{{\SetFigFont{5}{6.0}{\familydefault}{\mddefault}{\updefault}{\color[rgb]{0,0,0}i}%
}}}}
\put(1981,-7037){\makebox(0,0)[lb]{\smash{{\SetFigFont{5}{6.0}{\familydefault}{\mddefault}{\updefault}{\color[rgb]{0,0,0}j}%
}}}}
\put(3421,-7036){\makebox(0,0)[lb]{\smash{{\SetFigFont{5}{6.0}{\familydefault}{\mddefault}{\updefault}{\color[rgb]{0,0,0}i}%
}}}}
\put(4546,-7036){\makebox(0,0)[lb]{\smash{{\SetFigFont{5}{6.0}{\familydefault}{\mddefault}{\updefault}{\color[rgb]{0,0,0}j}%
}}}}
\put(181,-7081){\makebox(0,0)[lb]{\smash{{\SetFigFont{8}{9.6}{\familydefault}{\mddefault}{\updefault}{\color[rgb]{0,0,0}=\quad$\delta_{ij}$}%
}}}}
\put(2161,-6901){\makebox(0,0)[lb]{\smash{{\SetFigFont{5}{6.0}{\familydefault}{\mddefault}{\updefault}{\color[rgb]{0,0,0}$i\nu=0$}%
}}}}
\put(4726,-6901){\makebox(0,0)[lb]{\smash{{\SetFigFont{5}{6.0}{\familydefault}{\mddefault}{\updefault}{\color[rgb]{0,0,0}$i\nu=0$}%
}}}}
\end{picture}%
 
\label{Suszep}\notag
\end{eqnarray}
and obtain the N\'eel susceptibility $\chi^{zz}$ through
\begin{equation}
\chi^{zz}\big[{\bf q} = (\pi, \pi), \omega = 0\big] =  \sum_{j}(-1)^{j_x + j_y} \chi_{0j}^{zz}(i \nu = 0)\;,
\end{equation}
where $j_x$ $(j_y)$ denotes the distance in x(y)-direction to an arbitrarily chosen 
reference point. The transverse spin-spin correlation function $\chi_{ij}^{xx}$ 
is defined analogously.
These quantities only have a strict physical interpretation at $\Lambda = 0$, but we will consider them for finite $\Lambda$ as well and interpret a divergent flow as an indicator 
for a magnetic instability of the system. Again, using the physical temperature as a cutoff or working at $T>0$ would make this interpretation more straightforward, 
but would also strongly increase the numerical effort. The experience from other fermionic systems\cite{MetznerRMP} shows that divergences with respect to $\Lambda$ and $T$ are qualitatively equivalent.

\begin{figure}[t]
	\centering
	\includegraphics[width=0.95\linewidth]{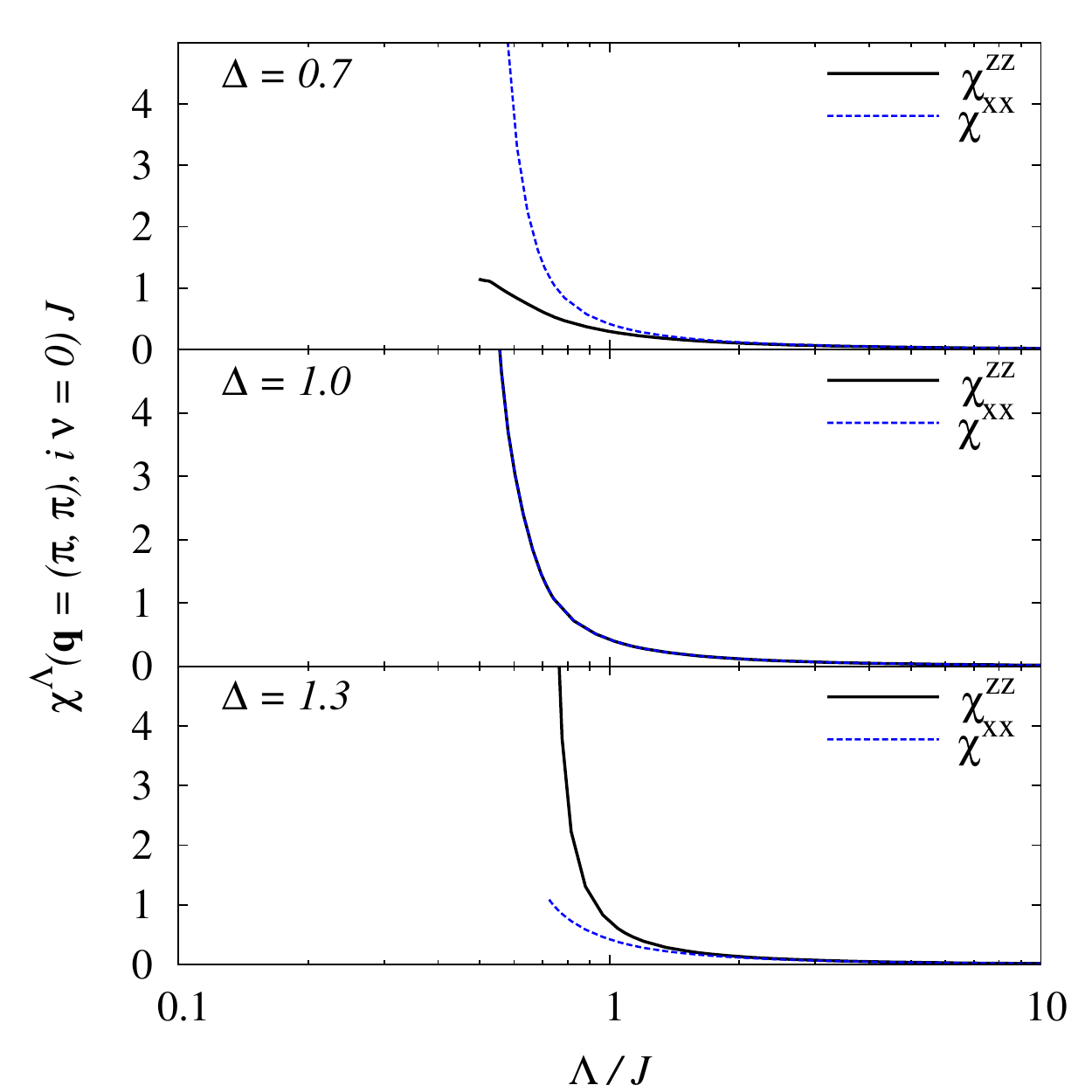}
	\caption{(Color online) Flow of the static transverse ($\chi^{xx}$) and longitudinal 
	($\chi^{zz}$) susceptibility for different anisotropy parameters $\Delta$. The 
	results confirm the existence of a phase transition between planar and axial 
	ordering at $\Delta = 1$.
\label{pic:Suszep}}
\end{figure}
The results for different anisotropies $\Delta$ are shown in Fig.~\ref{pic:Suszep}.
For $\Delta < 1$ we find a dominating transverse susceptibility $\chi^{xx}$, whereas for 
$\Delta > 1$ the longitudinal susceptibility $\chi^{zz}$ diverges at a larger scale $\Lambda$. 
At the isotropic point $\Delta = 1$ both susceptibilities are identical as required by the 
full $SU(2)$ invariance. This finding thus reproduces the well-known\cite{Farnell-04} phase transition 
between planar and axial ordering at $\Delta = 1$.
Furthermore, it has to be stressed that due to the generic divergence of the RG flow at a finite scale $\Lambda_c$ it is impossible to obtain
quantitative results for the susceptibility which could be compared to other methods on an absolute scale.

\subsection{Critical scales and temperatures}
\begin{figure}[t]
	\centering
	\includegraphics[width=0.95\linewidth]{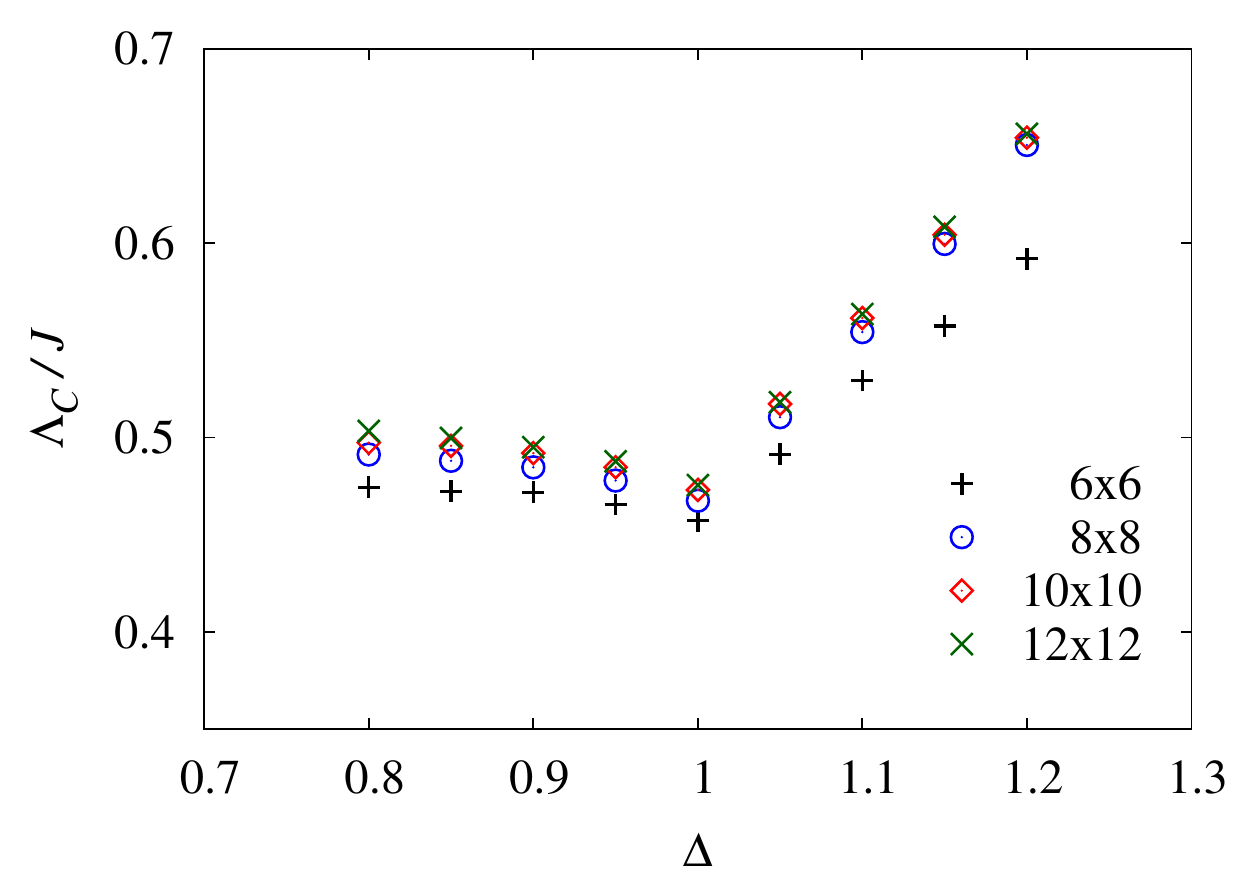}
	\caption{(Color online) Critical flow parameter $\Lambda_c$ extracted from the 
	divergence of the spin-fRG flow as a function of the 
	anisotropy parameter $\Delta$ for different system sizes. $\Lambda_c$ increases with the system size.}
	\label{pic:N}
\end{figure}

\begin{figure}[t]
	\centering
	\includegraphics[width=0.95\linewidth]{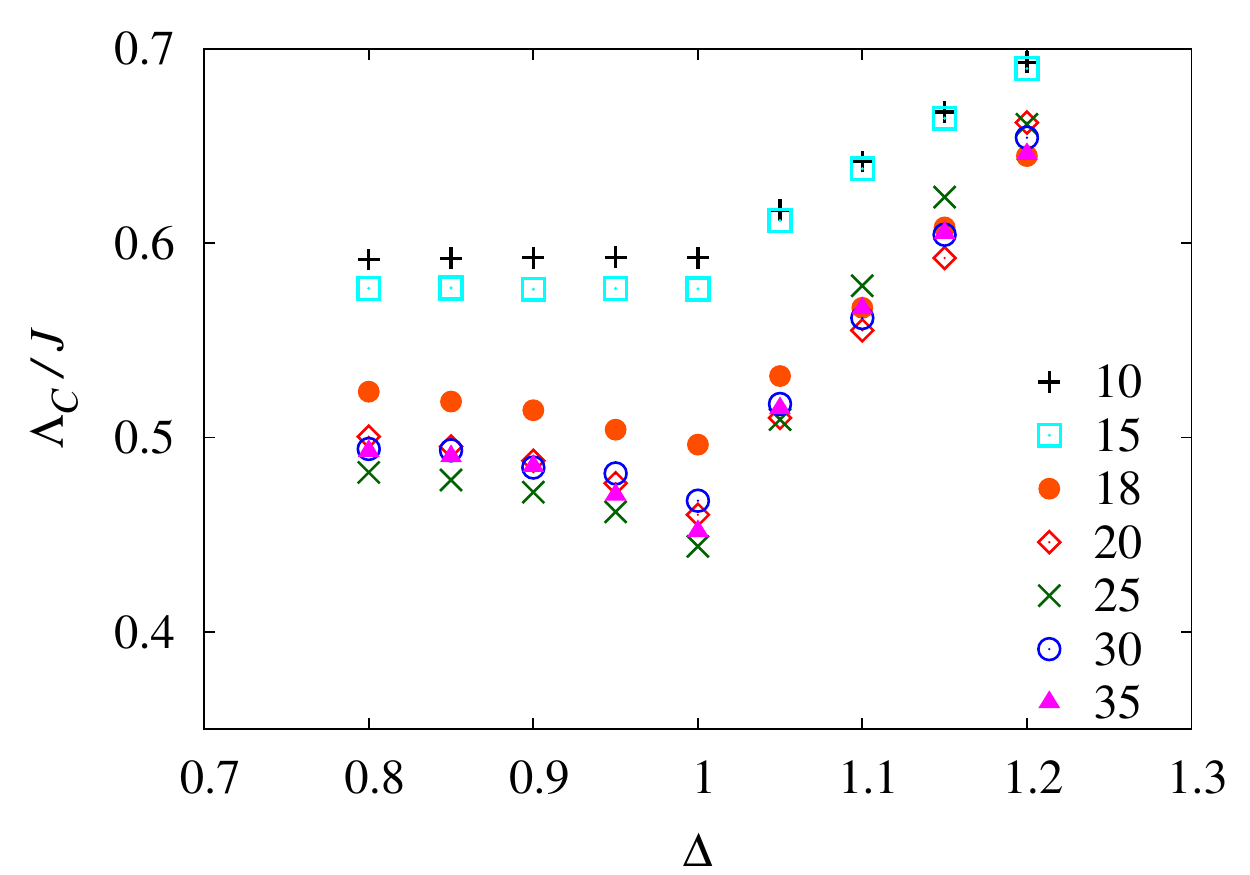}
	\caption{(Color online) Critical scale $\Lambda_c$ for different numbers of positive frequencies $N_f$ used in the discretization.}
	\label{pic:Fre}
\end{figure}

In Ref.~\onlinecite{Reuther-11} Reuther \textit{et al.} directly interpret the scale $\Lambda_c$, 
at which the fRG flow diverges, as a critical temperature $T_c$ for the phase transition into the 
ordered phase. In order to test this interpretation we have first studied the dependence
of $\Lambda_c$ on the system size and the frequency discretization. The results shown in Fig. \ref{pic:N} suggest that
the scale $\Lambda_c$ is indeed rather constant at least in the range of lattice sizes studied
here. A mild finite-size effect is, however, visible and indicates that $\Lambda_c$ tends to increase with system size.
As one can see in Fig. \ref{pic:Fre}, for a large enough number of frequencies $N_f$ the scale $\Lambda_c$ is essentially independent of the discretization.
In contrast to the system size dependence no monotonous dependence on $N_f$ is observed.
Still for larger $\Delta$, the relative difference in $\Lambda_c$ for different numbers of frequencies becomes smaller.

\begin{figure}[t]
	\centering
	\includegraphics[width=0.95\linewidth]{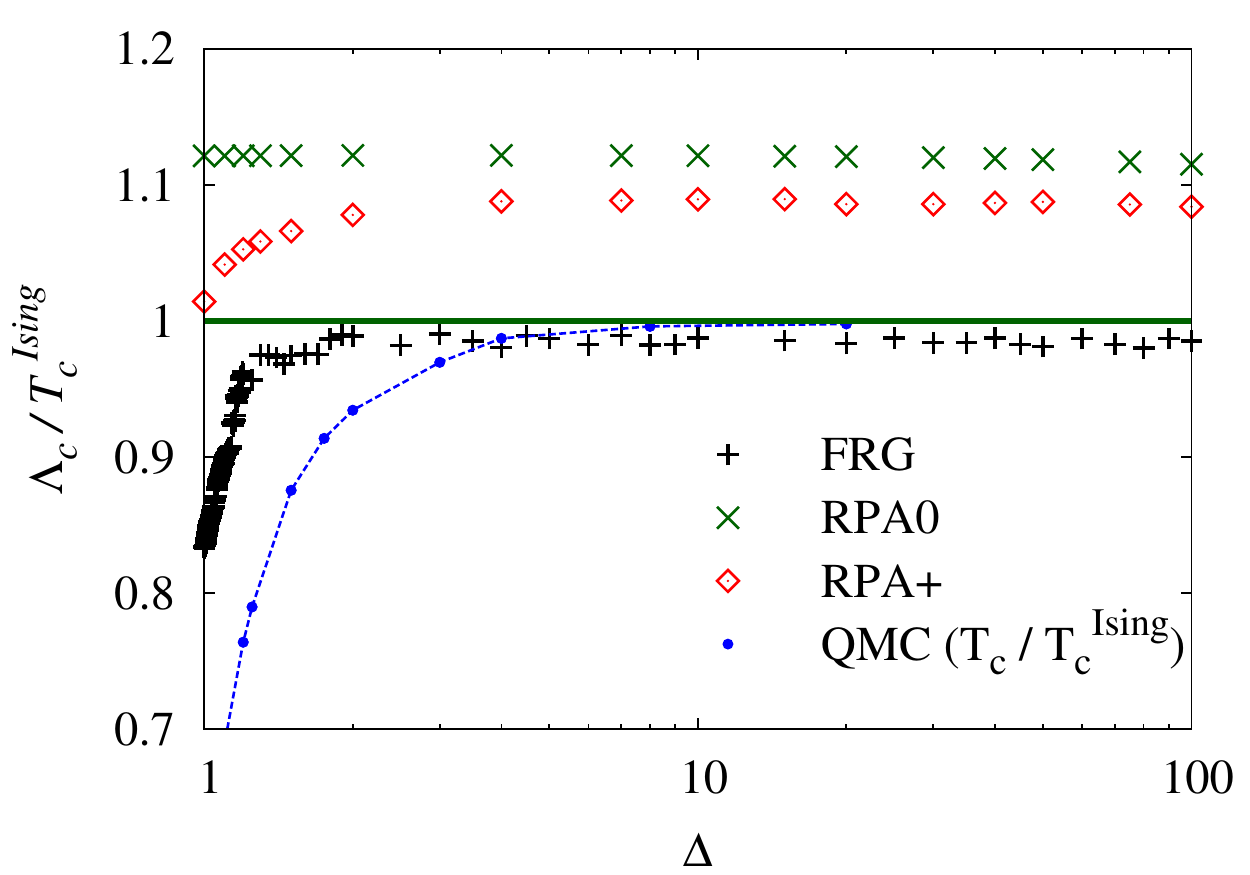}
	\caption{(Color online) Critical flow parameter $\Lambda_c$ as obtained from 
	different approaches in comparison to the exact result for the critical temperature
	in the Ising limit. In comparison to the RPA approaches the spin-fRG
	reproduces the Ising limit far more correctly.
	The errors of the QMC data are smaller than the size of the symbols.}
	\label{pic:Ising}
\end{figure}

As a next test we analyze the Ising limit ($\Delta \rightarrow \infty$) for which the 
scale $\Lambda_c$ is shown in Fig.~\ref{pic:Ising}. As discussed in Sec. \ref{sec:model}, 
the critical temperature $T_c = 2.269\Delta J/4$ is known exactly in this limit.
We observe that as a function of large $\Delta$ the values of $\Lambda_c$ of the 
RPA0, the RPA+ and the spin-fRG converge, but that the limiting values differ. 
Comparing the result for the RPA0 and RPA+ approximations we conclude that 
the inclusion of the self-energy improves the results, but that both methods still overestimate 
the ordering tendencies of the system. The spin-fRG presents an improvement over the RPA 
as the calculated $\Lambda_c$ agrees rather well with the critical temperature $T_c$ 
in the Ising limit. From this finding we may conclude that 
$\Lambda_c$ can indeed be interpreted as a critical temperature $\Lambda_c=T_c$ 
(at least in the Ising limit). However, as we show next, this interpretation becomes 
unsatisfactory when leaving the regime of strong Ising anisotropy.

\begin{figure}[t]
	\centering
	\includegraphics[width=0.95\linewidth]{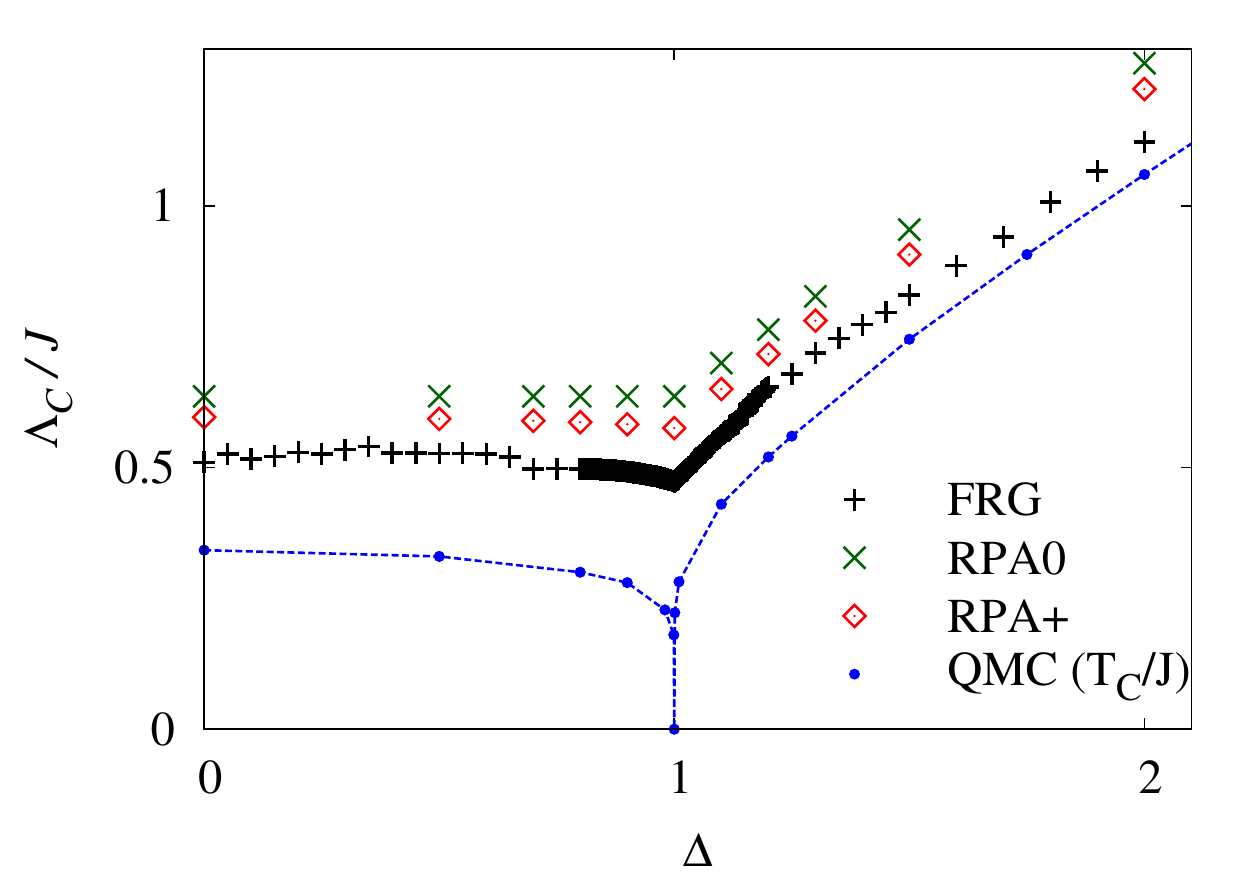}
	\caption{(Color online) Critical flow parameter $\Lambda_c$ as a function of the anisotropy 
	parameter $\Delta$ near the isotropic point. The errors of the QMC data are smaller 
	than the size of the symbols.}
	\label{pic:Tc}
\end{figure}

In particular, we consider the regime $\Delta\approx 1$ where we compare our results with the critical temperature
determined from the QMC analysis. As can be seen from Fig.~\ref{pic:Tc}, the spin-fRG 
improves the RPA in this regime as well, but the deviations from the QMC results are significantly 
larger than in the Ising limit. Nevertheless, except for the direct vicinity of the 
transition at $\Delta=1$, the spin-fRG scale $\Lambda_c$ qualitatively follows the dependence of $T_c$ 
on the anisotropy. For $\Delta \gtrsim 2.0$ the fRG critical scale $\Lambda_c$ even provides a 
reasonable quantitative estimate for $T_c$, whereas in the easy-plane regime fRG results deviate by roughly a factor of two from the QMC estimates. 
Thus except for the Ising limit the interpretation of the critical scale $\Lambda_c$ 
as a critical temperature $T_c$ has to be taken with great caution.

The strongest deviations between the critical scale extracted from the spin-fRG and the
QMC result are found close to the isotropic point $\Delta =1$, where the fRG 
yields a finite critical temperature in violation of the Mermin-Wagner theorem\cite{Mermin-66}.
Nonetheless the $\Delta$-dependence of $T_c$ obtained in the spin-fRG shows 
a kink-like feature at $\Delta = 1$, which is not observable in the RPA0 calculation 
and is only very weak in the RPA+. One may argue that at $T=0$ ordering indeed occurs and hence the nonzero 
critical scale obtained from spin-fRG does not violate the Mermin-Wagner theorem.
We note, however, that the isotropic Heisenberg model is gapless and it is not clear 
which energy scale $\Lambda_c$ thus represents. A more plausible way to understand 
the data is that the truncated spin-fRG fails to capture a significant part of the physics of collective 
fluctuations that would be needed to restore $T_c = 0$.

\subsection{Auxiliary fermion self-energy}
Closer inspection of the RPA data for $\Lambda_c$ reveals that RPA+ shows a
(weak) kink at $\Delta=1$ while the RPA0 data are monotonically increasing. The 
main difference between the two methods is the generated self-energy of the
auxiliary fermions. Thus a more detailed analysis of this quantity in the spin-fRG 
may lead to some insights.
 
\begin{figure}[t]
	\centering
	\includegraphics[width=0.95\linewidth]{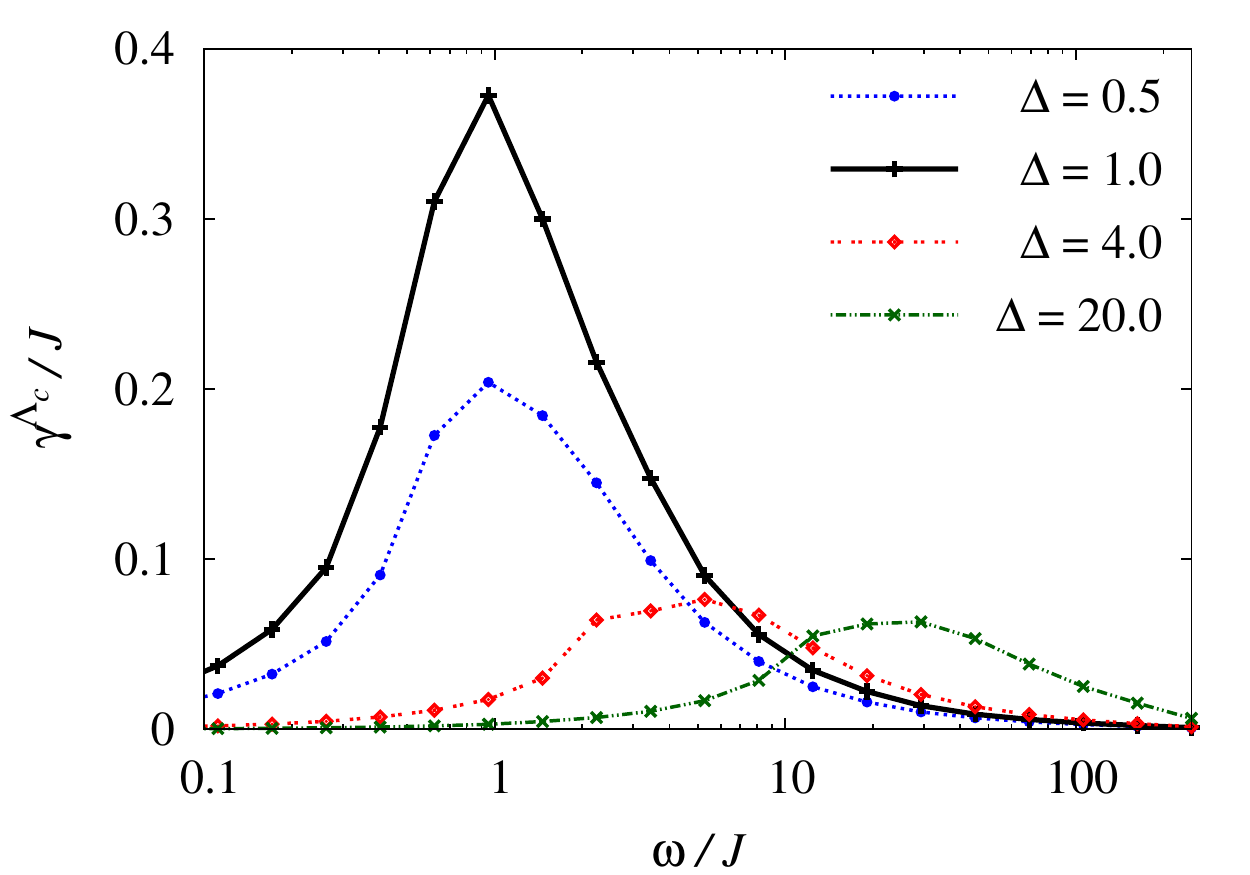}
	\caption{(Color online) Generated self-energy at the end of the fRG flow, i.e.,
	at $\Lambda=\Lambda_c$, for different values of $\Delta$. During the flow the
	peak builds up and moves in from higher frequencies.}
	\label{pic:self-energy} 
\end{figure}

The auxiliary fermion self-energy generated in the spin-fRG is not a 
physical quantity, since the auxiliary fermion Green's function connects 
physical and unphysical sectors of the fermionic Hilbert space. Nevertheless, 
analyzing the self-energy sheds some light on the shortcomings of the spin-fRG 
and the differences to other methods. At the end of the flow, near $\Lambda_c$, the 
frequency-dependent self-energy  $\gamma^\Lambda (\omega)$  shows a peak structure 
(see Fig. \ref{pic:self-energy}), where the maximum occurs at $\omega \approx \Delta J$ at least for $\Delta \geq 1$. 
In this regard, the fRG self-energy qualitatively differs from the one found in self-consistent 
RPA\cite{brinckmann}, which has a peak near zero frequency. The delayed generation 
of the self-energy in the fRG may be the reason why the self-consistent RPA of 
Ref.~\onlinecite{brinckmann} and the RPA+ implemented here do not agree although the 
same class of diagrams is kept in both approaches. 

For a two-site system the exact auxiliary fermion self-energy can be 
calculated\cite{Reuther-PhD} at vanishing chemical potential from a Lehmann representation 
yielding $\gamma \sim 1/\omega$. However, for this case the self-energy generated  
by the fRG possesses the same form as for the 2D model considered 
here (see Refs.~\onlinecite{Goettel-11,Kimmo-11}), i.e., with a peak at nonzero 
frequency whose position depends on $\Lambda$. This difference possibly originates 
from an incorrect description at small frequencies. 
Based on this hypothesis, we can give an argument why the spin-fRG critical scales 
agree with the QMC results for $T_c$ in the Ising limit. Here, the flow diverges already 
at larger scales $\Lambda$. As the self-energy enters the flow equations via the propagators 
$\sim 1/[\omega + \gamma(\omega)]$, its form for $\omega \ge \Lambda$ only weakly affects 
the flow. However, for $\Delta \approx 1$, the flow reaches smaller scales and this influence 
becomes more and more important. This qualitatively explains the good agreement in the 
Ising limit and may to some extent (more discussion is given further below) account for the 
failure of the method to correctly describe the critical temperature near the isotropic point
and, in particular, the violation of the Mermin-Wagner theorem. Here a stronger (or singular)
self-energy at small frequencies would help to avoid the divergence of the flow at a 
non-zero scale. We note that in self-consistent studies in the auxiliary fermion language
\cite{brinckmann} the results could be improved by an artificially introduced stronger
suppression of the auxiliary fermion spectral function at small frequencies. Similarly in the 
spin-fRG, a manipulation towards a larger self-energy at small frequencies would lead to a 
critical temperature $T_c = 0$ at the Heisenberg point and therefore the fulfillment of the Mermin-Wagner theorem,
but the critical temperature especially in the easy-plane regime would be strongly affected and vanishing, too.

\section{Discussion}

The goal of this paper was a comparison of spin-fRG results with those of other methods to 
perform a qualitative and quantitative assessment of the spin-fRG. We were able to confirm the 
phase transition between planar and axial ordering at the isotropic point $\Delta = 1$. 
The identification of the cutoff scale $\Lambda_c$ with the critical temperature $T_c$
yields rather accurate results in the Ising limit, but at the same time leads to a violation of 
the Mermin-Wagner theorem at the isotropic Heisenberg point and to deviations in the 
easy-axis and, to larger degree, in the easy-plane regime. The generated auxiliary 
fermion self-energy seems too weak to prevent divergences in the fRG flow for 
$\Lambda_c > 0$. Possible reasons for this error are as follows:

(i) The long-wave physics is not included as the system is still finite. We cannot exclude that the spin waves are just too constrained by our limited system size. 
We note, however, that for the available system sizes we observe an increasing rather than decreasing $\Lambda_c$ with system size (see Fig. \ref{pic:N}).

(ii) The mapping from the spin- to a fermionic system, where the constraint is fulfilled only on average and does not apply to quantum fluctuations. 
Here it appears problematic that the auxiliary fermion self-energy remains finite at small frequencies, in contrast with the exact solution of the the two-site problem, 
and possibly in contrast with the requirement of zero particle-number fluctuations of the auxiliary fermions.
The finiteness of the self-energy is a consequence of the structure of the frequency flow, and major changes to the formalism may be needed to alter this behavior.
This projection problem could also be investigated more systematically by including a nonzero temperature, where an exact projection scheme using an additional imaginary chemical potential 
is available\cite{Popov-88}. The major disadvantage consists of the fact that the imaginary chemical potential breaks various symmetries increasing the numerical effort substantially. 
For the Heisenberg model such calculations have been performed in Ref.~\onlinecite{Reuther-PhD}. As the results showed no qualitative differences to the average projection scheme, 
it seems that this is not the main source of error.

(iii) The RG treatment of the system with a truncation after the four-point vertex is not controlled 
because there is no small expansion parameter in the XXZ model. In bosonic descriptions of O(3)-models, the four-boson coupling is essential to drive the critical 
temperature to zero and hence to fulfill the Mermin-Wagner theorem. In fermionic language, this coupling would correspond to a 1PI eight-fermion vertex that is not present in the truncation 
used here for the spin-fRG. In this sense, the interaction between low-lying spin fluctuations that suppresses also the zero-frequency ordering might not be implemented correctly in this approach. 
This is also supported by the observation that the critical scales deviate more strongly from the QMC results for $T_c$ on the easy-plane side where the spin waves are gapless.

Further tests have been performed for one-dimensional models\cite{Goettel-11,Kimmo-11}. 
Here again the spin-fRG performs better than the RPA. For example in the latter there still
exists a finite scale $\Lambda_c$ at which the flow breaks down, while in the former the RG 
flow can be performed down to the smallest frequency $\omega_{min}$. The spin-fRG
correctly reproduces qualitative features like antiferromagnetic correlations.  
However, on a more quantitative level, the power-law decay of the spatial correlations 
in the Heisenberg chain is not reproduced by the spin-fRG, which rather predicts an incorrect 
exponential behavior. Similar shortcomings are seen in other models, e.g., the spin-fRG
generically yields too short-ranged correlations in both gapless and for gapped systems. 
Furthermore the ground-state energies and their dependence on the system parameters and
number of lattice sites are not reproduced. Hence, the spin-fRG cannot be regarded as a 
reliable theoretical method for the study of one-dimensional spin system.

\section{Conclusion}
In conclusion, the spin-fRG considerably improves RPA results for low-dimensional quantum 
spin systems. The qualitative behavior of the spin correlations on short scales is reproduced 
correctly. For the two-dimensional XXZ model studied here, the spin-fRG does detect the 
quantum phase transition at $\Delta=1$ and the critical scale $\Lambda_c$ can be used as an 
estimate for ordering temperatures away from the isotropic point. 
For all anisotropies $|\Delta -1|\gtrsim 0.1$, the deviations in $\Lambda_c$ from $T_c$ are less than a factor of 2.
At the isotropic point $\Delta =1$, however, the spin-fRG predicts a finite ordering temperature
in violation of the Mermin-Wagner theorem, which also spoils the quantitative validity of the
spin-fRG results in the vicinity of $\Delta=1$. Furthermore, as discussed above, due to the 
generic divergence of the RG flow at a finite scale $\Lambda_c$ it is impossible to obtain
quantitative results for the susceptibilities or correlation functions which could be compared to other methods 
on an absolute scale. These two drawbacks are well known from the fRG treatment of itinerant two-dimensional many-fermion systems\cite{metznerRMP}, 
and also the appealingly advanced approximation level of the spin-fRG with the full-self-energy feedback and frequency-dependence cannot remedy this deficiency. 
Hence, one may conclude that the spin-fRG should be rather used to explore phase diagrams 
in situations where strongly differing types of ground states are expected to compete, 
i.e., ground states that differ in their correlations already on short distances. 
Such situations were actually addressed in most of the spin-fRG papers up to now.\\

\acknowledgments
We thank J. Reuther for the helpful exchange on several technical questions, and 
W. Brenig, L. Fritz, A. Gendiar, V. Meden, W. Metzner, 
H. Schoeller, and R. Thomale for useful discussions.
We particularly thank also K. S\"a\"askilahti for collaboration on the 
application of the spin-fRG to spin chains.
This work was supported by the German Research Foundation (DFG) 
through FOR 723, FOR 912, and the Emmy-Noether Program (D.S.).


\end{document}